\documentclass{Interspeech}

\interspeechcameraready

\title{Fine-Tuning Text-to-Speech Diffusion Models Using Reinforcement Learning with Human Feedback}

\author[affiliation={1,2}]{Jingyi}{Chen}
\author[affiliation={2}]{Ju Seung}{Byun}
\author[affiliation={1}]{Micha}{Elsner}
\author[affiliation={3}]{Pichao}{Wang}
\author[affiliation={2}]{Andrew}{Perrault}

\affiliation{Department of Linguistics}{The Ohio State University}{USA}
\affiliation{Department of Computer Science and Engineering}{The Ohio State University}{USA}
\affiliation{}{Amazon}{USA}

\email{{chen.9220, byun.83, elsner.14, perrault.17}@osu.edu, pichaowang@gmail.com}

\usepackage{comment}
\usepackage{comment}
\usepackage{subfigure} 
\usepackage{url}
\usepackage{hyperref}
\usepackage[utf8]{inputenc} 
\usepackage[T1]{fontenc}    
\usepackage{booktabs}       
\usepackage{amsfonts}       
\usepackage{nicefrac}       
\usepackage{microtype}      
\usepackage{xcolor}         
\usepackage{amsmath}

\usepackage{graphicx}
\usepackage{amssymb}
\usepackage[noend]{algpseudocode}
\usepackage{algorithm}
\usepackage{lipsum}
\newcommand{\pretrained}{p_\mathrm{pre}}

\hypersetup{hidelinks,
colorlinks=true,
linkcolor=black,
urlcolor=black,
citecolor=black}
\usepackage{caption}
\usepackage{xurl}

\begin{document}

\maketitle

\footnotetext[3]{This work does not relate to the author's position at Amazon.}

\begin{abstract}
Diffusion models produce high-fidelity speech but are inefficient for real-time use due to long denoising steps and challenges in modeling intonation and rhythm. To improve this, we propose Diffusion Loss-Guided Policy Optimization (DLPO), an RLHF framework for TTS diffusion models. DLPO integrates the original training loss into the reward function, preserving generative capabilities while reducing inefficiencies. Using naturalness scores as feedback, DLPO aligns reward optimization with the diffusion model’s structure, improving speech quality. We evaluate DLPO on WaveGrad 2, a non-autoregressive diffusion-based TTS model. Results show significant improvements in objective metrics (UTMOS 3.65, NISQA 4.02) and subjective evaluations, with DLPO audio preferred 67\% of the time. These findings demonstrate DLPO’s potential for efficient, high-quality diffusion TTS in real-time, resource-limited settings. 
\end{abstract}

\section{Introduction}
Text-to-speech (TTS) synthesis is a fundamental technology driving applications ranging from virtual assistants to accessibility tools. However, building effective TTS aims to achieve low latency, operate within limited computational resources, and maintain high-quality output. Meeting these stringent requirements demands the development of efficient generative models capable of balancing quality and speed.

Diffusion probabilistic models have recently emerged as powerful generative models, achieving remarkable success in tasks such as image and video synthesis~\cite{ho2022video,ho2020denoising,nichol2021improved,yang2023diffusion}. These models excel at capturing complex, high-dimensional data distributions through iterative refinement, making them a compelling choice for TTS synthesis. In this domain, diffusion models have demonstrated significant potential to generate high-quality, natural-sounding speech. However, their adoption in practical scenarios remains challenging for several reasons. First, diffusion models are inherently computationally intensive due to their iterative denoising process, which conflicts with the efficiency demands. Second, TTS synthesis requires the sequential generation of audio waveforms, introducing unique challenges such as ensuring temporal coherence and acoustic consistency across thousands of samples. Third, diffusion models often struggle to model subtle yet essential aspects of speech, such as intonation, rhythm, and emotional expressiveness, leading to outputs that can sound unnatural or inconsistent. However, these challenges may stem not from inherent limitations of diffusion models but rather from factors such as insufficient model scaling, suboptimal training objectives, and the need for improved human preference alignment.

These challenges highlight the need for innovative optimization techniques to make diffusion-based TTS models both efficient and effective. Recent advances in reinforcement learning with human feedback (RLHF) have shown promise in improving generative models by aligning them with human preferences. For example, techniques like reward-weighted regression (RWR)~\cite{peters2007reinforcement}, denoising diffusion policy optimization (DDPO)\footnote{Black et al. (2023). Training diffusion models with reinforcement learning. arXiv:2305.13301.}, and diffusion policy optimization with KL regularization (DPOK)~\cite{fan2024reinforcement} have been successfully applied to text-to-image synthesis. These methods leverage human feedback to guide optimization, enhancing the quality and relevance of generated outputs. However, directly applying these techniques to TTS diffusion models is challenging. TTS tasks not only demand greater temporal coherence and acoustic precision but also involve evaluation metrics like naturalness (e.g., MOS scores) and intelligibility (e.g., WER), which are less straightforward to optimize compared to image quality metrics. RLHF enhances speech quality and naturalness in the VoiceCraft transformer TTS model~\cite{peng2024voicecraft}, as shown by recent work \footnote{Chen et al. (2024). Enhancing Zero-shot Text-to-Speech Synthesis with Human Feedback. arXiv:2406.00654}, While recent work (Nagaram, 2024) \footnote{Nagaram, R. (2024). Enhancing Emotional Expression in Text-to-Speech Models through Reinforcement Learning with AI Feedback. *Technical Report No. UCB/EECS-2024-23*} has applied RL to enhance emotion expression in Grad-TTS \cite{popov2021grad}; however, issues related to speech quality and naturalness persist.

To improve TTS diffusion models, we introduce Diffusion Loss-Guided Policy Optimization (DLPO), a novel RLHF framework designed for this task. DLPO introduces a unique reward regularization strategy by directly integrating the diffusion model’s original training loss into the reward function. This innovation addresses the key limitations of existing RLHF methods and serves two critical purposes: 1) preserve the model's inherent capabilities: by aligning the reward function with the original diffusion training objective, DLPO ensures that the model maintains its ability to generate high-quality speech while adapting to human feedback;
2) prevent overfitting and deviation: the original diffusion loss acts as a stabilizing regularizer, balancing external reward optimization with the preservation of the model`s probabilistic structure, thereby mitigating over-optimization issues commonly observed in TTS.

We evaluate DLPO on a reproduction of WaveGrad 2, a non-autoregressive diffusion-based TTS model specifically designed for efficient waveform generation. WaveGrad 2~\cite{chen2021wavegrad} eliminates the sequential dependencies of autoregressive models, significantly reducing inference latency. While diffusion models typically struggle with real-time deployment, its streamlined architecture provides an ideal testbed for studying reinforcement learning techniques, allowing us to focus on the core synthesis process without the added complexity of hybrid systems. Additionally, our approach enhances efficiency, making them more suitable for applications. Experimental results demonstrate that DLPO significantly improves both objective and subjective metrics. DLPO achieves a UTMOS score of 3.65 and a NISQA score of 4.02, while maintaining a low WER of 1.2. In human evaluations, DLPO-generated audio is preferred in 67\% of pairwise comparisons with model before fine-tuning, underscoring its ability to enhance naturalness and acoustic quality while aligning with the efficiency.

These findings suggest that DLPO offers a robust framework for fine-tuning TTS diffusion models. Beyond TTS, its integration of task-specific regularization with reinforcement learning provides a promising approach to improving generative models in other sequential and time-sensitive domains. We provide demo audios in \href{https://demopagea.github.io/DLPO-demo/}{https://demopagea.github.io/DLPO-demo/} and the code in \href{https://anonymous.4open.science/r/DLPO-6556/}{https://anonymous.4open.science/r/DLPO-6556/}

\section{Methods}
\begin{figure}[t]
  \centering
  \includegraphics[width=0.8\linewidth]{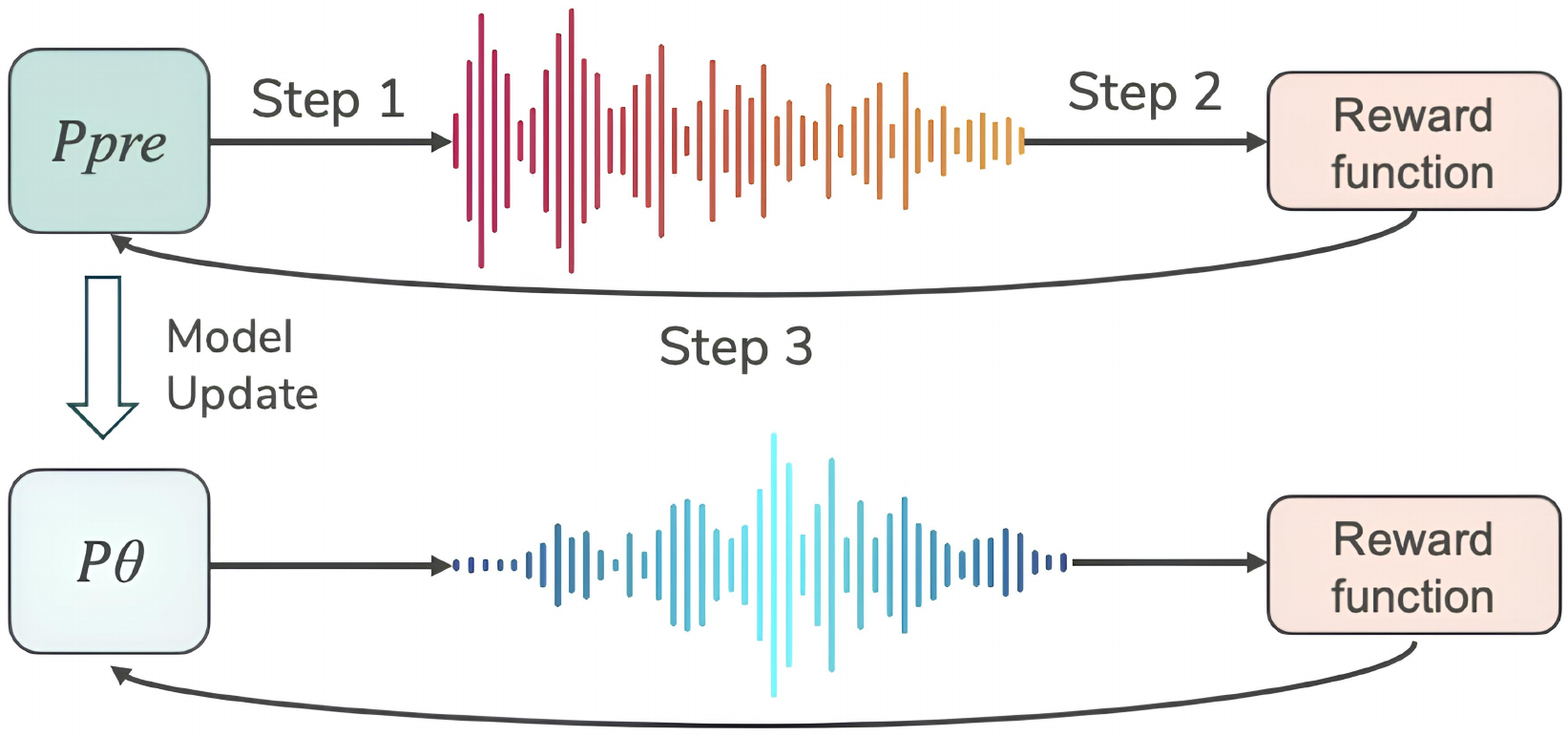}
  \caption{Procedure for fine-tuning diffusion TTS with RLHF}
  \label{fig:rl}
\end{figure}

Our methodology leverages RLHF to fine-tune a pretrained TTS model. To simulate human feedback, we employ reward predictions from the UTokyo-SaruLab mean opinion score (UTMOS) prediction system, enabling objective evaluation of naturalness, intelligibility, and human preference. As shown in \autoref{fig:rl}, the pretrained model, \( P_{\text{pre}} \), generates an audio sample. The reward model, UTMOS \cite{saeki2022utmos}, evaluates the naturalness of the generated speech and assigns a reward. This reward is then used to update the model from \( P_{\text{pre}} \) to \( P_\theta \). The same procedure is repeated to further update \( P_\theta \).

We conduct experiments by applying text-to-image RL algorithms, including RWR, DDPO, DPOK, and KLinR, to fine-tune the base model, alongside our proposed Diffusion Loss-guided Policy Optimization (DLPO) method. We trained for 5.5 hours on 8 A100-SXM-80GB GPUs with a batch size of 64 and 10 denoising steps. By integrating the diffusion model loss into the reward function, DLPO improves synthesized speech quality while efficiently preventing the model from deviating excessively from the base model. 

\subsection{Models}
\label{WaveGrad 2}
In this study, we use the PyTorch reproduction of WaveGrad 2 provided by MINDs Lab (\href{https://github.com/maum-ai/WaveGrad 2}{https://github.com/maum-ai/WaveGrad 2}) as the pretrained TTS diffusion model---we refer to this specific model as WaveGrad 2R. WaveGrad 2 is a non-autoregressive TTS model that implements diffusion denoising probabilistic model (DDPM)~\cite{ho2020denoising}. WaveGrad 2 models the conditional distribution $p_\theta(y_0 |x)$ where $y_0$ represents the waveform and $x$ the associated context. The distribution follows the reverse of a Markovian forward process $q(y_t | y_{t-1})$, which iteratively introduces noise to the data.

Reversing the forward process can be accomplished by training a neural network $\mu_\theta(x_t,c,t)$ with the following objective:
\begin{align}
\label{eqddpm}
\mathcal{L}_{DDPM}(\theta) = & \, \mathbb{E}_{c \sim p(c)} \mathbb{E}_{t \sim \mathcal{U}\{0, T\}} \mathbb{E}_{p_\theta(x_{0:T} | c)} \Big[ \Vert \tilde{\mu}(x_t, t) \notag \\
& \quad - \mu_\theta(x_t, c, t) \Vert_2 \Big]
\end{align}
where $\tilde{\mu}$ is the posterior mean of the forward process and $x_t$ is the prediction at timestep $t$ in the denoising process. This objective is justified as maximizing a variational lower bound on the log-likelihood of the data, which is trained to predict the scaled derivative by minimizing the distance between the ground truth added noise $\epsilon$ and the model prediction:
\begin{align}
    \mathbb{E}_{c \sim p(c)} \mathbb{E}_{t\sim \mathcal{U}\{0,T\}} \mathbb{E}_{p_\theta(x_{0:T}|c)} \left[\Vert \tilde{\epsilon}(x_t,t) - \epsilon_\theta(x_t,c,t)\Vert_2\right]
\end{align}
where $\tilde{\epsilon}(x_t,t)$ is the ground truth added noise for step $x_t$ and $\epsilon_\theta(x_t,c,t)$ is the predicted noise  for step $x_t$. 

Sampling from a diffusion model begins with drawing $x_T \sim \mathcal{N}(0,I)$ and following the reverse process $p_\theta(x_{t-1} |x_t,c)$ to produce a trajectory \{$x_T,x_{T-1},...,x_0$\} ending with a sample $x_0$. 
As discussed in~\cite{chen2020wavegrad,chen2021wavegrad}, the linear variance schedule used in~\cite{ho2020denoising} is adapted in the sampling process of WaveGrad 2, where the variance of the noise added at each step increases linearly over a fixed number of timesteps. The linear variance schedule is a predefined function that dictates the noise variance added at each step of the forward diffusion process. This schedule is crucial because it determines how much noise is added to the data at each timestep, which affects the quality of the generated samples during the reverse denoising process. 

\subsubsection{Reward model}
\label{reward}
We use the UTMOS prediction system~\cite{saeki2022utmos} to predict the speech quality of generated audios from WaveGrad 2R. Mean opinion score (MOS) is a subjective scoring system that allows human evaluators to rate the perceived quality of synthesized speech on a scale from 1 to 5 (the greater MOS, the better speech quality of the generated audios), which is one of the most commonly employed evaluation methods for TTS system~\cite{streijl2016mean}. UTMOS is trained to capture nuanced audio features that reflect human judgments of speech quality, offering a reliable MOS prediction without requiring extensive labeled data. Thus, we use UTMOS as the reward model for fine-tuning WaveGrad 2R. UTMOS is trained on datasets from the VoiceMOS Challenge 2022~\cite{huang2022voicemos}, including 14 hours of audio from male and female speakers with MOS ratings.

\subsubsection{RL for fine-tuning TTS diffusion models }

We model denoising as a $T$-step finite horizon Markov decision process (MDP). Defined by the tuple $(S, A, \rho_0, P, R)$, an MDP consists of a state space $S$, an action space $A$, an initial state distribution $\rho_0$, a transition kernel $P$, and a reward function $R$. At each timestep $t$, an agent observes a state $s_t$ from $S$, selects an action $a_t$ from $A$, receives a reward $R(s_t, a_t)$, and transitions to a new state $s_{t+1} \sim P(s_{t+1} | s_t, a_t)$. The agent follows a policy $\pi_\theta(a|s)$, parameterized by $\theta$, to make decisions. As the agent operates within the MDP, it generates trajectories, sequences of states and actions $(s_0, a_0, s_1, a_1, . . . , s_T , a_T )$.

The goal of reinforcement learning (RL) is to maximize the expected total reward across trajectories produced under its policy. We define the objective function as $ \mathcal{J}_{\mathrm{RL}}(\theta) = \mathbb{E}_{\pi_\theta} \left[\sum_{t=0}^T R(\mathrm{s_t},\mathrm{a_t})\right]$, where $\mathcal{J}_{\text{RL}}(\theta)$ is the expected total reward function, and the expectation is taken over the trajectories generated by the policy $\pi_\theta$.

   

We formulate a MDP for the denoising phase of WaveGrad 2R, where the model iteratively refines noisy signals to generate cleaner, more natural speech.
\begin{equation}
\label{eq1}
\begin{aligned}
    s_t &\triangleq (c, x_{T-t}) , \quad 
    a_t \triangleq x_{T-t-1}, \quad \\
    P(s_{t+1} | s_t, a_t) &\triangleq (\delta_c, \delta_{a_t}) \\ 
    \rho(s_0) &\triangleq (p(c), \mathcal{N}(0, I)) \\
    R(s_t, a_t) &\triangleq 
    \begin{cases}  
         r(s_{t+1}) = r(x_0, c), & \text{if } t = T-1,  \\  
         0, & \text{otherwise,} 
    \end{cases}
\end{aligned}
\end{equation}
where $\delta$ is the Dirac delta distribution, $c$ is the text prompt sampled from $p(c)$, and $r(x_0,c)$ is the reward model UTMOS introduced in \autoref{reward}. $s_t$ and $a_t$ are the state and action at timestep $t$, $\rho(s_0)$ and $P$ are the initial state distribution and the dynamics, and $R$ is the reward function. We let $\pi_\theta(a_t|s_t)\triangleq p_\theta(x_{T-t-1}|x_{T-t},c)$ be the initial parameterized policy, which is the pretrained WaveGrad 2R. Trajectories consist of $T$ timesteps, after which $P$ leads to a terminating state. The cumulative reward of each trajectory is equal to $r(x_0,c)$. Thus, optimizing policy $\pi_\theta$ with RL to maximize $\mathcal{J}_{\mathrm{DDRL}}$ is equivalent to fine-tuning WaveGrad 2 to maximize the UTMOS of generated speech:
\begin{align}
\label{ddrl}
    \mathcal{J}_{\mathrm{DDRL}}(\theta) = \mathbb{E}_{c \sim p(c)}\mathbb{E}_{x_0 \sim p_\theta (x_0|c)} \left[ r(\mathrm{{x_0},\mathrm{c}})\right]    
\end{align}

\subsection{ Diffusion loss policy optimization (DLPO)}
We propose enhancing the RL training process by incorporating the diffusion model objective into the reward function as a penalty. Previous work by \cite{ouyang2022training} has demonstrated that mixing pretraining gradients with RL gradients leads to improved performance on various NLP tasks, outperforming a reward-only approach. Following this insight, we hypothesize that adding the diffusion model loss to the objective function can similarly boost performance and prevent model degradation, particularly in maintaining temporal coherence. Specifically, we propose the following objective:
\begin{equation} \label{eq10} \mathbb{E}_{c \sim p(c)}\mathbb{E}_{p_\theta(x_{0:T}|c)} \left[-\alpha r(x_0,c)-\beta\Vert \tilde{\epsilon}(x_t,t) - \epsilon_\theta(x_t,c,t)\Vert_2 \right] \end{equation}
where $\alpha$ and $\beta$ represent the reward and the weight of the diffusion model loss, respectively. We extend the work of \cite{ahmadian2024back} by incorporating the diffusion model objective as a penalty in the reward function. This formulation aligns with the training procedure of TTS diffusion models, where adding the diffusion model penalty $\beta \Vert \tilde{\epsilon}(x_t,t) - \epsilon_\theta(x_t,c,t)\Vert_2$ to the reward function prevents deviations from the model's intended output and improves diffusion TTS naturalness. 
\subsection{Experiments and results}
WaveGrad 2R is pre-trained on the LJSpeech dataset~\cite{ljspeech17} which consists of 13,100 short audio clips of a female speaker and the corresponding texts, totaling approximately 24 hours. We use WaveGrad 2R's training set and validation set to fine-tune WaveGrad 2R (12388 samples for training, 512 samples for validation). 200 unseen samples are used as test set for evaluation.

For automatic evaluation of the fine-tuned models, we use another pretrained speech quality and naturalness assessment model (NISQA)~\cite{mittag2021nisqa}, trained on the NISQA Corpus including more than 14,000 speech samples produced by male and female speakers along with samples' MOS ratings. (This use of a separate MOS network is intended to guard against overfitting the reward model.) We also conduct a human evaluation on the speech quality of the generated audios from fine-tuned WaveGrad 2R. To evaluate the intelligibility of the synthesized audio, we transcribe the speech with a pre-trained ASR model, Whisper~\cite{radford2023robust}, and compute the word error rate (WER) between the transcribed text and original transcript.

We save the top three checkpoints for each model during training and use them to generate audios for 200 unseen texts. We then use UTMOS and NISQA to predict MOS score of these generated audios and the real human speech audios, labeled as ground truth.

\begin{table*}[htbp]
\centering
\footnotesize
\caption{Comparison of diffusion model RL fine-tuning methods}
\label{tab:diffusion_model_methods}
\begin{tabular}{|l|p{10.5cm}|l|l|l|}  
\hline
\textbf{Method} & \textbf{RL objective} & \textbf{UTMOS$ \uparrow$} & \textbf{NISQA $\uparrow$} & \textbf{WER $\downarrow$} \\ \hline
\textbf{Ground truth} & N/A & 4.20 & 4.37 & 0.99\% \\\hline
\textbf{WaveGrad 2R} & N/A & 2.90 & 3.74 & 1.5\% \\\hline
\textbf{RWR} & $\mathbb{E}_{p(c)}\mathbb{E}_{\pretrained(x_0|c)} \left[- r(x_0,c)\log p_\theta(x_0|c)\right]$ & 2.18 & 3.00 & 8.9\% \\\hline
\textbf{DDPO} & 
\(\mathbb{E}_{p(c)}\mathbb{E}_{p_\theta(x_{0:T}|c)} 
\left[\sum_{t=1}^T - \log p_\theta(x_{t-1}|x_t,c)r(x_0,c)\right]\) & 2.69 & 2.96 & 2.1\% \\\hline
\textbf{DPOK} & 
\(\begin{aligned}
\mathbb{E}_{p(c)}[\alpha\mathbb{E}_{p_\theta(x_{t-1}|x_t,c)} &\left[- r(x_0,c) \log p_\theta(x_{t-1}|x_t,c)\right] + \\ &\beta \sum_{t=1}^T \mathbb{E}_{p_\theta (x_t|c)} \left[\mathrm{KL}(p_\theta(x_{t-1}|x_t,c) \Vert \pretrained(x_{t-1}|x_t,c))\right]]
\end{aligned}\) & 3.18 & 3.76 & \textbf{1.1}\% \\\hline

\textbf{KLinR} & 
\(\mathbb{E}_{p(c)}\mathbb{E}_{p_\theta(x_{0:T}|c)} 
\big[- \left( r(x_0,c)- \mathrm{KL}(p_\theta(x_0|c)\Vert \pretrained(x_0|c)\right) \
 \sum_{t=1}^T \log p_\theta(x_{t-1}|x_t,c)\big]\) & 3.02 & 3.73 & 1.3\% \\\hline
\textbf{DLPO} & 
\(\mathbb{E}_{p(c)} \mathbb{E}_{p_\theta(x_{1:T}|c)} 
\Big[- (\alpha r(x_0,c)-\beta\nabla_\theta\Vert \tilde{\epsilon}(x_t,t)  -  \epsilon_\theta(x_t,c,t)\Vert_2) 
 \log p_\theta(x_{t-1}|x_t,c)\big]\) & \textbf{3.65} & \textbf{4.02} & 1.2\% \\\hline
\textbf{OnlyDL} & 
\(\mathbb{E}_{p(c)} \mathbb{E}_{p_\theta(x_{1:T}|c)} 
\Big[(\beta\nabla_\theta\Vert \tilde{\epsilon}(x_t,t)  -  \epsilon_\theta(x_t,c,t)\Vert_2) 
 \log p_\theta(x_{t-1}|x_t,c)\big]\) & 3.16 & 3.45 & 1.4\% \\\hline
\end{tabular}
\end{table*}

\subsubsection{RL algorithms and results}
We compare RL algorithms in fine-tuning WaveGrad 2R. Results are shown in \autoref{tab:diffusion_model_methods}. 
Reward-weighted regression (RWR)~\cite{peters2007reinforcement} is proposed to optimize the diffusion denoising loss $\mathcal{L}_{DDPM}$ weighted by a reward function $r(x_0, c)$.
We find that RWR degrades speech quality in fine-tuning TTS diffusion models, leading to noisy audios. The training reward UTMOS decreases to 2.18, the evaluation NISQA MOS to 3.00 and WER worsens to 8.9\%.
This deterioration might be due to the model being fine-tuned on a static dataset produced by a pre-trained model, and RWR relies on an approximate log-likelihood by disregarding the sequential aspect of the denoising process and only using the final samples $x_0$. The denoising loss $\mathcal{L}_{DDPM}$ \autoref{eqddpm} does not compute an exact log-likelihood; it is instead a variational bound on $\log p_\theta (x_0 | c)$. As such, the RWR procedure approach to training diffusion models lacks theoretical justification and only approximates optimizes $\mathcal{J}_{\mathrm{DDRL}}$.

Differing from RWR, the denoising diffusion policy optimization (DDPO) aims to directly optimize $\mathcal{J}_{DDRL}$ using the score function policy gradient estimator, also known as REINFORCE~\cite{williams1992simple,mohamed2020monte}. DDPO alternately collects denoising trajectories ${x_T , x_{T-1} , . . . , x_0 }$ via sampling and updates parameters based on these denoising trajectories via gradient descent. This estimator only allows for one step of optimization for each data collection round, since the gradient needs to be calculated using the entire denoising trajectories derived from the current parameters. Although DDPO is shown to achieve better performance in fine-tuning the text-to-image diffusion model, it fails to improve TTS diffusion models on speech quality. The training reward UTMOS decreases to 2.69, the evaluation NISQA MOS decreases to 2.96 and WER is 2.1\%. The failure of DDPO to improve TTS diffusion models in terms of speech quality can be attributed to the inherent limitations of its optimization process. DDPO only allows for one step of optimization for each data collection round, requiring gradients to be computed across entire denoising trajectories derived from the current parameters. While this approach has been successful in fine-tuning text-to-image diffusion models, it is less effective for TTS models due to the increased complexity of generating sequential audio waveforms in the time domain. Audio waveforms demand a higher degree of temporal coherence and structural fidelity compared to static images, making single-step optimization insufficient for capturing the intricate dependencies required for high-quality speech synthesis.

Diffusion Policy Optimization with a KL-shaped Reward (DPOK) was proposed to address the deterioration problem with RWR. DPOK adds KL regularization between the fine-tuned and pre-trained models $KL(p_\theta(x_0|z)\Vert \pretrained(x_0|z) $ to the objective function helps to mitigate overfitting of the diffusion models to the reward and prevents excessively diminishing the "skill" of the original diffusion model. Our results show that DPOK improves WaveGrad 2R by increasing the reward UTMOS from 2.9 to 3.18 during training and the evaluation NISQA from 3.74 to 3.76. WER decreases to 1.1\%. 
We also implement KLinR~\cite{ahmadian2024back}, another RL objective which leverages a KL penalty to prevent degradation in the coherence of the model. In contrast to DPOK, this objective function $\mathcal{J}_{\mathrm{DDRL}}(\theta)$ includes the KL penalty within the reward function. KLinR also presents some improvement in WaveGrad 2R by increasing the reward UTMOS from 2.9 to 3.02 during training and WER decreases to 1.1\%, but NISQA is slightly lower than the base WaveGrad 2R model, which is 3.73.

These results show that adding KL regularization can help stabilize the model but the improvement of WaveGrad 2R is limited. The limited improvement of DPOK and KLinR in TTS diffusion models, despite their success in text-to-image models, can be attributed to fundamental differences between the tasks and the alignment of these regularization techniques with the TTS fine-tuning process. TTS operates in the time domain, requiring the sequential generation of audio waveforms with temporal coherence and acoustic consistency, which makes it more sensitive to inconsistencies that KL regularization may not fully address. While the KL penalty stabilizes training by mitigating overfitting and preserving coherence, it may inadvertently limit the exploration of task-specific innovations necessary for significant performance gains in TTS, particularly for perceptual quality, naturalness, and intelligibility. 

Unlike DPOK and KLinR, which rely on KL divergence to stabilize training, DLPO directly leverages the original diffusion model loss to maintain the model's inherent capabilities while allowing for greater flexibility in adapting to task-specific requirements.  Experimental results demonstrate that DLPO outperforms DPOK and KLinR in training UTMOS increasing to 3.65 and NISQA scores increasing to 4.02. DLPO has WER 1.2\%, which is similar to DPOK and KLinR. 

To further investigate the impact of using the original TTS diffusion model loss as a regularizer, we conducted an experiment where the baseline WaveGrad 2R was fine-tuned using only the diffusion loss as the reward, referred to as OnlyDL. The loss function is outlined in \autoref{tab:diffusion_model_methods}. We trained this model for 5.5 hours, matching the training time of DLPO.
During training, both the total loss and diffusion loss plateaued and exhibited minimal fluctuations throughout training. Additionally, the UTMOS for OnlyDL remained consistently around 3.16 during the entire training period, whereas DLPO demonstrated a steady UTMOS improvement from approximately 3 to above 3.65. OnlyDL has NISQA as 3.45 and maintained a low WER of 1.4\%. These results suggest that using the original TTS diffusion model loss as a regularizer effectively prevents model deviation and preserves the inherent capabilities of the TTS diffusion model. However, combining the original TTS diffusion model loss with reward-based UTMOS optimization is crucial for enhancing the naturalness of the TTS diffusion model while preventing deviation.



\noindent \textbf{Human evaluation.} We further conduct an experiment recruiting 11 human participants to evaluate the quality of speech generated by a previous version of DLPO-tuned WaveGrad 2R vs.\ base WaveGrad 2R. We randomly selected 20 pairs of outputs from 200 unseen texts. 
Evaluators were asked to assess which audio had better speech naturalness and quality. In 67\% of comparisons, audios generated from DLPO fine-tuned WaveGrad 2R were rated as better than audios generated by the base WaveGrad 2R model, while 14\% of comparisons have audios generated by the baseline WaveGrad 2R model rated as better. 19\% of comparisons were rated as about the same. Using a binomial test, the DLPO-tuned model is better with $p<10^{-16}$.

\section{Discussion}
This work tackles the challenges of fine-tuning TTS diffusion models using reinforcement learning techniques. While existing methods like RWR and DDPO struggle to address the unique temporal and acoustic demands of TTS, we propose Diffusion Loss-Guided Policy Optimization (DLPO) as a tailored solution. Integrating the diffusion model’s original training loss into the reward function, DLPO stabilizes the training, prevents overfitting, and enables task-specific adaptations.
DLPO effectively overcomes the limitations of prior RLHF methods by aligning reward optimization with the probabilistic structure of diffusion models, facilitating better modeling of temporal dependencies, prosody, and speech nuances. This approach demonstrates the importance of leveraging task-specific regularization to address the complexities of sequential data generation. Our findings establish DLPO as a robust framework for advancing diffusion-based TTS synthesis and set a foundation for broader applications in resource-constrained and real-time scenarios. While DLPO enhances speech naturalness, future work will focus on expanding language support, refining prosody control, and further improving speaker adaptability to increase its versatility.


\section{Acknowledgements}
We are thankful for the generous support of computational resources provided by the Ohio Supercomputer Center.

\bibliographystyle{IEEEtran}
\bibliography{mybib}

\end{document}